\documentclass[sigconf, screen=true]{acmart}

\usepackage{booktabs} 
\usepackage{balance}
\graphicspath{ {Figures/} }
\usepackage{tabularx}
\usepackage{float}
\usepackage{amsmath}
\usepackage{nameref}
\usepackage{enumitem}

\usepackage{tikz}
\usetikzlibrary{calc}
\usetikzlibrary{trees}
\usetikzlibrary{arrows}
\usetikzlibrary{graphs}
\usetikzlibrary{positioning}
\usetikzlibrary{shapes.geometric}
\usetikzlibrary{shapes.misc}
\tikzstyle{block}=[draw opacity=0.7,line width=1.0cm]

\tikzset{
modal/.style={>=stealth',shorten >=1pt,shorten <=1pt,auto,node distance=1.5cm, semithick},
world/.style={circle,draw,minimum size=0.5cm,fill=gray!15},
ghost/.style={circle,draw,minimum size=0.5cm,fill=white,draw=white, fill opacity=0},
atom/.style={rectangle,minimum size=0.5cm,fill=green!50!black!20,draw=green!50!black},
structure/.style={=>stealth’,shorten >=1pt,shorten <=1pt,auto,node distance=25mm, thick, minimum size=5mm},
set/.style={rectangle, minimum size=0.5cm,fill=green!50!black!20,draw=green!50!black, font=\sffamily\scriptsize},
lattice/.style={=>stealth’,shorten >=1pt,shorten <=1pt, auto, node distance=1cm, semithick, minimum size=3mm},
every node/.style={font=\sffamily\footnotesize},
point/.style={circle,draw,inner sep=0.5mm,fill=black},
scriptsize/.style={font=\sffamily\scriptsize},
tiny/.style={font=\sffamily\tiny},
highlight/.style={fill=red!50!black!20,draw=red!50!black},
comment/.style={rectangle, fill=white,text opacity=1,fill opacity=0, draw opacity=0, minimum size=0.5cm,font=\itshape\sffamily\tiny},
reflexive above/.style={->,loop,looseness=7,in=120,out=60},
reflexive below/.style={->,loop,looseness=7,in=240,out=300},
reflexive left/.style={->,loop,looseness=7,in=150,out=210},
reflexive right/.style={->,loop,looseness=7,in=30,out=330}
}

\DeclareMathOperator{\rsquare}{\mathrel{\square}}

\DeclareMathOperator{\requires}{\mathrel{requires}}
\DeclareMathOperator{\require}{\mathrel{require}}
\DeclareMathOperator{\refines}{\mathrel{refines}}

\DeclareMathOperator{\contains}{\mathrel{contains}}
\DeclareMathOperator{\conflicts}{\mathrel{conflicts}}

\DeclareMathOperator{\equals}{\mathrel{equals}}
\DeclareMathOperator{\satisfies}{\mathrel{satisfies}}
\DeclareMathOperator{\satisfy}{\mathrel{satisfy}}

\usepackage{color}

\newcommand{\T}{Tarski}
\newcommand{\E}{ECAS}
\newcommand{\case}{Electronically Controlled Air Suspension (ECAS) System}
\newcommand{\type}[1]{\textsc{#1}}

\setcopyright{acmcopyright}
\acmPrice{15.00}
\acmDOI{10.1145/3106237.3122825}
\acmYear{2017}
\copyrightyear{2017}
\acmISBN{978-1-4503-5105-8/17/09}
\acmConference[ESEC/FSE'17]{2017 11th Joint Meeting of the European Software Engineering Conference and the ACM SIGSOFT Symposium on the Foundations of Software Engineering}{September 4--8, 2017}{Paderborn, Germany}

\begin{document}
\title[A Tool for Automated Reasoning about Traces Based on \\ Configurable Formal Semantics]{A Tool for Automated Reasoning about Traces Based on Configurable Formal Semantics}

\author{Ferhat Erata}
\affiliation{%
  \streetaddress{Wageningen University, the Netherlands}
}
\email{ferhat@computer.org}

\author{Arda Goknil}
\affiliation{%
  \streetaddress{University of Luxembourg, Luxembourg}
}
\email{arda.goknil@uni.lu}

\author{Bedir Tekinerdogan}
\affiliation{%
  \streetaddress{Wageningen University, the Netherlands}
}
\email{bedir.tekinerdogan@wur.nl}

\author{Geylani Kardas}
\affiliation{
  \streetaddress{Ege University, Turkey}
}
\email{geylani.kardas@ege.edu.tr}

\renewcommand{\shortauthors}{F. Erata, A. Goknil, B. Tekinerdogan and G. Kardas}

\begin{abstract}
We present Tarski, a tool for specifying configurable trace semantics to facilitate automated reasoning about traces. Software development projects require that various types of traces be modeled between and within development artifacts. For any given artifact (e.g., requirements, architecture models and source code), Tarski allows the user to specify new trace types and their configurable semantics, while, using the semantics, it automatically infers new traces based on existing traces provided by the user, and checks the consistency of traces. It has been evaluated on three industrial case studies in the automotive domain (\url{https://modelwriter.github.io/Tarski/}).
\end{abstract}

%
%

\ccsdesc[500]{Software and its engineering~Consistency}
\ccsdesc[500]{Software and its engineering~Traceability}
\ccsdesc[300]{Software and its engineering~Specification languages}
\ccsdesc[300]{Software and its engineering~Formal methods}

\keywords{Traceability; Domain-Specific Modeling; Formal Trace Semantics; Automated Reasoning; Alloy; KodKod}

\maketitle

\section{Introduction} 
\label{sec:intro}


The complexity of software systems in safety critical domains (e.g. avionics and automotive) has significantly increased over the years. Development of such systems requires various phases which result in several artifacts (e.g., requirements documents, architecture models and test cases). In this context, traceability~\cite{Ramesh:2001aa, SWEBOK:IEEEComputerSociety:2014} not only establishes and maintains consistency between these artifacts but also helps guarantee that each requirement is fulfilled by the source code and test cases properly cover all requirements, a very important objective in safety critical systems and the standards they need to comply with~\cite{DO178C, ISO26262}. As a result, the engineers have to establish and maintain several types of traces, having different semantics, between and within various development artifacts. 

We present a tool, \T \footnote{The name is inspired by Alfred Tarski's foundational work on the relational calculus}, which supports specifying configurable trace semantics to facilitate multiple forms of automated trace reasoning. \T~is developed for environments, requiring maintenance of various artifacts, within the context of our research~\cite{ModelWriter, Assume} in collaboration with Ford-Otosan~\cite{FordOtosan}, Airbus~\cite{Airbus} and Havelsan~\cite{Havelsan}. The motivation behind \T~is to provide a way to interactively specify trace types and semantics, which vary for different artifacts, to be used in automated trace reasoning. 

There are approaches and tools~\cite{TraceabilityOperationalSemantics2005, FormalTraceSemantics2014, ChangeImpact2014, SemanticTrace2011, GoknilKM13} that use a predetermined set of possible trace types and their semantics for automated reasoning. However, in the case of dealing with complex software systems, instead of a one-size-fits-all approach, it is required to enable the adoption of several trace types and their semantics, and herewith the various forms of automated reasoning about traces. To do so, \T~provides the following features: (i) specifying trace semantics which can be configured due to project and artifact types, (ii) deducing new traces based on the given trace semantics and on the traces which the engineer has already specified, and (iii) identifying the traces whose existence causes a contradiction according to the trace semantics. 
The tool provides a traceability domain model which describes the basic concepts of traceability such as \textit{Trace-link} and \textit{Trace-location}. The notion of trace-location refers to traceable elements in an artifact, while the notion of trace-link refers to traces between trace-locations. The user defines new trace types by extending \textit{Trace-link} and \textit{Trace-location}. The user specifies the semantics of new trace types in a restricted form of Alloy~\cite{AlloyBook}, i.e., First-Order Logic (FOL) augmented with the operators of the relational calculus~\cite{tarski1941calculus}. 
We employ Kodkod~\cite{Torlak:2007, KodKodPhdThesis}, an efficient SAT-based constraint solver for FOL with relational algebra and partial models, for automated trace reasoning using the trace semantics. Our tool is integrated with Eclipse~\cite{Eclipse} platform. 

In the remaining sections, we outline \T's features and components. We highlight the findings from our evaluation of \T~over multiple industrial case studies with one of our industrial partners.

\section{Related Work}
\label{sec:related}

Several approaches and tools have been proposed for automated trace reasoning using the trace semantics~\cite{TraceabilityOperationalSemantics2005, EgyedG05, Egyed03, EgyedG02, Cleland-HuangCC03, ConsistencyCheckingModelMerging, lamb2011, GoknilKB10, SemanticTrace2011, ChangeImpact2014, hove2009change, goknil2008change, FormalTraceSemantics2014, goknil2008metamodeling, Drivalos2008, DetectRepairConsistency2008}. These approaches employ a predefined set of trace types and their corresponding semantics. For instance, Goknil et al.~\cite{SemanticTrace2011} provide a tool for inferencing and consistency checking of traces between requirements using a set of trace types (e.g., \textit{refines}, \textit{requires}, and \textit{contains}) and their formal semantics. Similarly, Egyed and Gr{\"{u}}nbacher~\cite{EgyedG05} propose a trace generation approach. They do not allow the user to introduce new trace types and their semantics for automated reasoning. In the development of complex systems, it is required to enable the adoption of various trace types, and herewith automated reasoning using their semantics. 

\T~does not encode any predefined trace type or semantics. It allows the user to interactively define new trace types with their semantics to be used in automated reasoning about traces. Using the semantics specified by the user, \T~deduces new traces and checks the consistency of traces. 

\section{Tool Overview} \label{sec:overview}

\begin{figure}[ht]
\centering
\vspace*{-1.2em}
\includegraphics[width=0.95\linewidth]{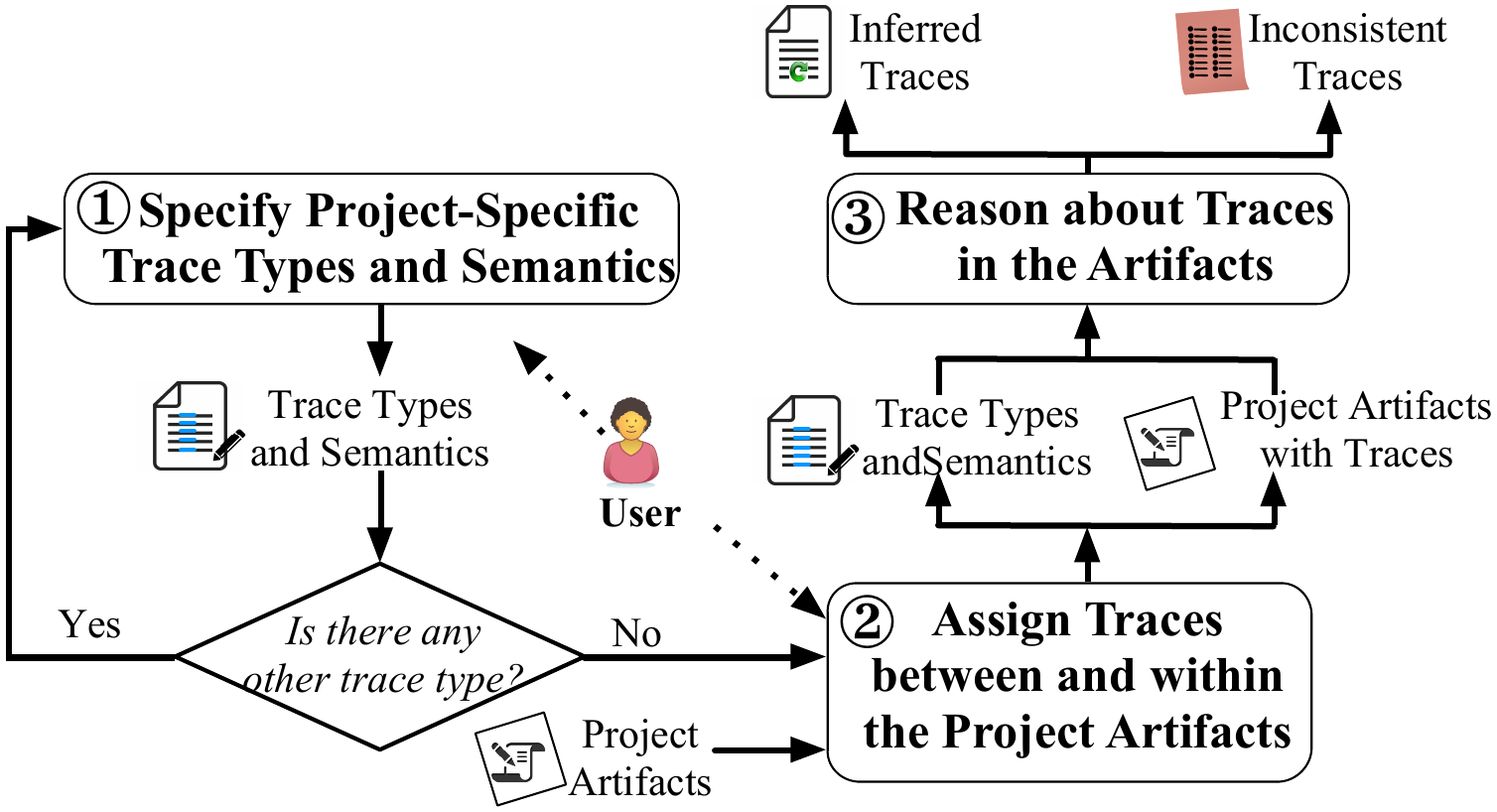}
\vspace*{-1.3em}
\caption{Tool Overview}\label{fig:tool_overview}
\vspace*{-1.5em}
\end{figure}

\T~is the tool supporting our approach for automated reasoning about traces based on configurable trace semantics, recently described in ~\cite{ferhat2017SAC:PL}. Fig.~\ref{fig:tool_overview} presents an overview of our tool. In Step 1, the user specifies new trace types and their semantics in First-Order Logic (FOL) augmented with the operators of the relational calculus~\cite{tarski1941calculus}. To do so, \T~employs a restricted form of Alloy~\cite{AlloyBook} with a custom text editor. New trace types are defined by extending \textit{Trace-link} and \textit{Trace-location} in \textit{Traceability Domain Model}. 

Once the user specifies the trace types and their semantics, \T~ allows the user to assign traces between and within the input project artifacts (e.g., requirements specifications, architecture models, and test cases) using the trace types (Step 2).  
After the traces are manually assigned, the tool proceeds to Step 3 with automated trace reasoning. In the rest of the section, we elaborate each step in Fig.~\ref{fig:tool_overview} using the Electronically Controlled Air Suspension (ECAS) System of Ford-Otosan~\cite{FordOtosan}, a safety-critical system in automotive domain, as a case study.   

\subsection{Specification of Trace Types and Semantics} \label{subsec:specification}
\begin{figure}[ht]
\centering
\includegraphics[width=\columnwidth]{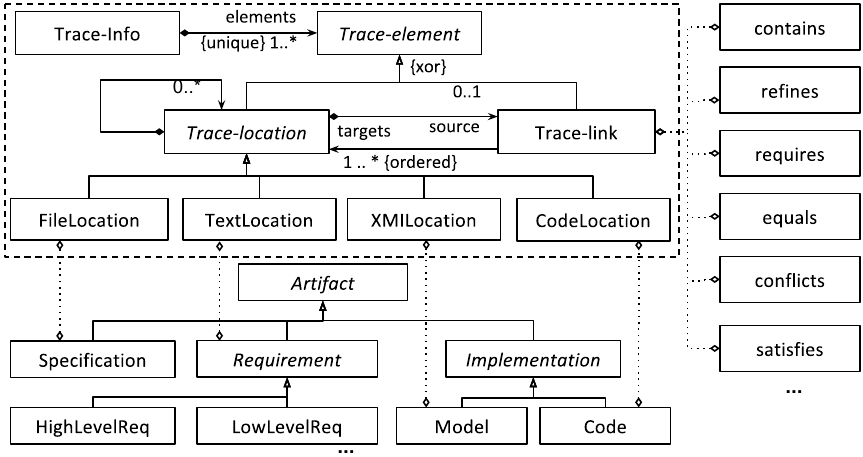}
\vspace*{-2.0em}
\caption{Traceability Model with User-defined Trace Types}
\label{fig:traceability_domain_model}
\vspace*{-1.5em}
\end{figure}

As the first step, for the artifacts, the user specifies trace types and their semantics in FOL using a restricted form of Alloy. First, the user extends the \textit{traceability domain model} with new trace and artifact types. Fig.~\ref{fig:traceability_domain_model} shows part of the extended traceability domain model for the \E~case study.

\begin{figure}[ht]
\centering
\vspace*{-1.5em}
\includegraphics[width=\columnwidth]{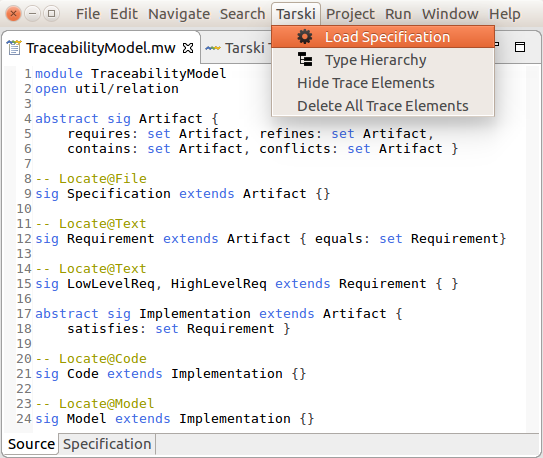}
\vspace*{-2.4em}
\caption{Some Example Trace Types in \T}
\label{fig:specification_types}
\end{figure}

We extend \textit{Trace-link} in Fig.~\ref{fig:traceability_domain_model} with new trace types (e.g., \textit{contains}, \textit{refines}, and \textit{satisfies}), while \textit{Text-location} is extended with new types of elements of the artifacts to be traced in the case study (e.g., \textit{Requirement}, \textit{HighLevelReq}, and \textit{Code}). Fig.~\ref{fig:specification_types} shows some of the extensions of \textit{Trace-link} and \textit{Trace-location} in Fig.~\ref{fig:traceability_domain_model}. 

In the following, we briefly explain the restricted Alloy notation \T~employs for declaring trace types and their semantics. 
Signatures define the vocabulary of a model in Alloy (see keyword \textit{sig} in Fig.~\ref{fig:specification_types}). We use them to extend \textit{Trace-location} for declaring artifact element types (see Lines 4, 9, 12, 15, 17, 21 and 24 in Fig.~\ref{fig:specification_types}). \T~employs some special annotations to specify artifacts' location types (Lines 8, 11, 14, 20 and 23). The location type information is later used by \T~to create the Eclipse workspace fields to save traces assigned in Step 2 in Fig.~\ref{fig:tool_overview} (see Section~\ref{subsec:assignment}). For instance, \textit{Requirement} is given as a text location in Line 11 (see \textit{Requirement} and \textit{Text-location} in Fig.~\ref{fig:traceability_domain_model}), while \textit{Code} is given as a source code location in Line 20. For a trace between a textual requirement and a code fragment, using the location information in Fig.~\ref{fig:specification_types}, \T~ creates a \textit{resource} field as a path referring to the location of the requirement, while the \textit{resource}, \textit{offset}, and \textit{length} fields are created to refer to the code fragment where \textit{resource} gives the path of the source code file, \textit{offset} gives the start index of the code fragment in the code file, and \textit{length} gives the length of the code fragment. 

\begin{figure}[ht]
\centering
\vspace*{-1.0em}
\includegraphics[width=\columnwidth]{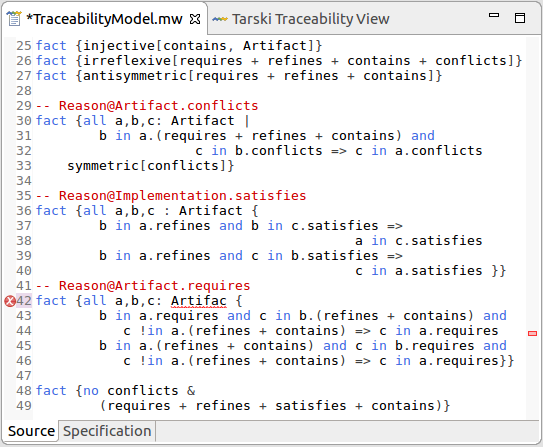}
\vspace*{-2.4em}
\caption{Example Trace Semantics in \T}
\label{fig:semantics}
\vspace*{-1.0em}
\end{figure}

New trace types are defined as binary relations in the signature fields (see Lines 5, 6, 12, and 18 in Fig.~\ref{fig:specification_types}). \T~automatically extends \textit{Trace-link} for those binary relations (see Fig.~\ref{fig:traceability_domain_model}). For instance, in Line 18, \textit{Satisfies} is declared as a new trace type between \textit{Implementation} and \textit{Requirement}. Trace semantics is given as \textit{facts} in Alloy (see Fig.~\ref{fig:semantics}). Facts are constraints that are assumed to always hold. They are used as axioms in constructing examples and counterexamples~\cite{AlloyBook}. 
The \textit{Refines}, \textit{Requires} and \textit{Contains} trace types are defined \textit{irreflexive} and \textit{antisymmetric} (see Lines 26 and 27 in Fig.~\ref{fig:semantics}). In addition, \textit{Contains} is \textit{injective} (Line 25).   

As part of the semantics, we define how trace types are related to each other (Lines 30-49). For instance, according to the fact in Lines 30-33 where \textit{a}, \textit{b} and \textit{c} are artifact elements, if \textit{a} \textit{refines}, \textit{requires} or \textit{contains} \textit{b}, while \textit{b} \textit{conflicts} with \textit{c}, then \textit{a} also \textit{conflicts} with \textit{c}.   

\subsection{Trace Assignment in Project Artifacts} \label{subsec:assignment}

\begin{figure}[ht] 
\centering
\vspace*{-1.0em}
\includegraphics[width=\columnwidth]{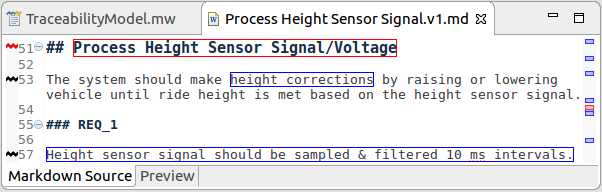}
\vspace*{-2.4em}
\caption{Part of the \E~Requirements Specification} \label{fig:annotations}
\vspace*{-1.0em}
\end{figure}

\T~guides the user in assigning traces between and within the input artifacts (see Step 2 in Fig.~\ref{fig:tool_overview}). The user manually assigns traces for the input artifacts using the trace types. The main challenge is that the traceable parts of textual artifacts (e.g., requirements in a requirements specification) need to be determined before assigning traces. To address this challenge, \T~employs a semantic parsing approach~\cite{Gyawali2017} that automatically maps natural language to Description Logic (DL) axioms. The mappings between the natural text and the DL axioms are used by \T~to automatically identify the traceable parts of textual artifacts. Fig.~\ref{fig:annotations} shows part of the \E~requirements specification after semantic parsing in \T.


The blue colour indicates the traceable parts of the \E~requirements specification which do not yet have any trace. When the user wants to assign a trace from/to these blue coloured parts, \T~automatically suggests the possible trace types using the type hierarchy encoded in Step 1 (see Fig.~\ref{fig:specification_types}). After the trace is assigned, the blue colour automatically becomes red, which indicates having at least one trace.


\subsection{Automated Reasoning about Traces} \label{subsec:reasoning}
Inferencing and consistency checking aim at deriving new traces based on given traces and determining contradictions among traces. 
These two activities enrich the set of traces in the artifacts. 
They are processed in parallel because the consistency checking uses the machinery for inferencing and also checks the inconsistencies among inferred traces as well as among given traces.

\begin{table}[ht]
\vspace*{-0.8em}
\caption{Some Requirements and Code Fragments in \E} \label{tab:artifacts}
\vspace*{-1.0em}
   \begin{tabular}{l p{0.87\columnwidth}}
    \toprule
    Nr. & Requirements/Code Fragments\\
    \midrule
    $r_{11}$ &The system shall do height corrections using long and short term filtered height sensor signal.\\   
    $r_{59}$ &The system shall always use height sensors in the range of 0-5V to avoid long term signal filtering.\\
    $r_{60}$ &The system shall do height corrections using long and short term filtered height sensor signal with 10ms interval.\\
    $r_{97}$ &The system shall filter height sensor signal in short term and long term for height corrections.\\
    $r_{98}$ &The system shall filter height sensor signal in long term for height corrections.\\
    \midrule
    $i_{14}$ &\texttt{vehicle::ecas::processHeightSensor::filterSignal}\\
    $i_{72}$ &\texttt{vehicle::ecas::processHeightSensor}\\
    \bottomrule
 \end{tabular}
\vspace*{-1.0em}
\end{table}

\subsubsection{Inferring New Traces} 
\label{subsubsec:inference}
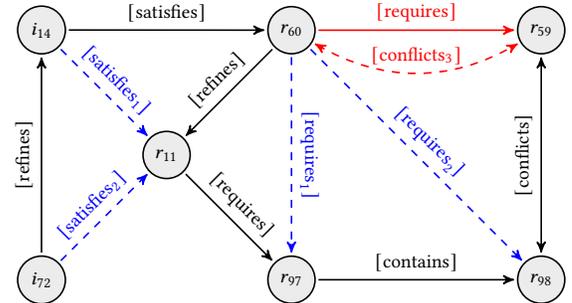
\begin{figure}[ht]
\centering
\vspace*{-1.0em}
    \begin{tikzpicture}[modal,node distance=1.7cm,world/.append style={minimum size=0.6cm}]
        \node[world] (11) [] {$r_{11}$};
        \node[world] (60) [above right=of 11] {$r_{60}$};
        \node[ghost] (60_59) [below right=of 60] {$r_{60}$};
        \node[world] (97) [below right=of 11] {$r_{97}$};
        \node[world] (59) [above right=of 60_59] {$r_{59}$};
        \node[world] (98) [below right=of 60_59] {$r_{98}$};
        
        \node[world] (14)[above left=of 11] {$i_{14}$};
        \path[->] (14) edge[] node[sloped,anchor=south] {$[\satisfies]$}  (60);
        \path[->] (14) edge[dashed, color=blue] node[sloped,anchor=south] {$[\satisfies_1]$}(11);

        \node[world] (72)[below left=of 11] {$i_{72}$};
        \path[->] (72) edge[] node[sloped,anchor=south] {$[\refines]$}  (14);
        \path[->] (72) edge[dashed, color=blue] node[sloped,anchor=south] {$[\satisfies_2]$}(11);

        \path[->] (60) edge[] node[sloped,anchor=south] {$[\refines]$}  (11);
        \path[->] (60) edge[dashed,color=blue] node[sloped,anchor=south] {$[\requires_1]$} (97);
        \path[->] (11) edge[] node[sloped,anchor=south]{$[\requires]$} (97);
        \path[->] (60) edge[color=red] node[sloped,anchor=south] {$[\requires]$}(59);
        \path[->] (60) edge[dashed, color=blue] node[sloped,anchor=south] {$[\requires_2]$}(98);
        \path[<->] (60) edge[bend right, dashed, color=red] node[sloped,anchor=south] {$[\conflicts_3]$}(59);
        \path[->] (97) edge[] node[sloped,anchor=south] {$[\contains]$}(98);
        \path[<->] (98) edge[] node[sloped,anchor=south] {$[\conflicts]$}(59);
    \end{tikzpicture}
\vspace*{-1.2em}
\caption{Example Inferred and Inconsistent Traces in \E} \label{fig:consistency_checking}
\vspace*{-1.2em}
\end{figure}

\T~ takes the artifacts and their manually assigned traces as input, and automatically deduces, using the user-defined trace types and their semantics, new traces as output. Fig.~\ref{fig:consistency_checking} gives the assigned and inferred traces for some simplified \E~requirements and source code fragments in Table~\ref{tab:artifacts}.   


The solid arrows represent the manually assigned traces, while the dashed arrows are the traces automatically inferred by \T. For instance, the user assigns the \textit{refines} traces between $i_{72}$ and $i_{14}$, and between $r_{60}$ and $r_{11}$. Using the trace semantics in Fig.~\ref{fig:semantics}, \T~automatically infers two \textit{satisfies} traces, two \textit{requires} traces and one \textit{conflicts} trace in Fig.~\ref{fig:consistency_checking}. For instance, $i_{14}$ \textit{satisfies} $r_{11}$ (i.e., inferred) because it \textit{satisfies} $r_{60}$ which \textit{refines} $r_{11}$ (see the \textit{fact} in Lines 36-40 in Fig.~\ref{fig:semantics}). The \textit{conflicts} trace between $r_{60}$ and $r_{59}$ is inferred because $r_{60}$ \textit{requires} $r_{98}$ which \textit{conflicts} with $r_{59}$ (see the \textit{fact} in Lines 30-33 in Fig.~\ref{fig:semantics}). Please note that the \textit{requires} trace between $r_{60}$ and $r_{98}$ is inferred.

\subsubsection{Checking Consistency of Traces.} \label{subsubsec:consistency}
\T~takes the artifacts and their given and inferred traces as input, and automatically determines, using the user-defined trace types and their semantics, the inconsistent traces as output. \T~provides an explanation of inconsistent traces by giving all the manually assigned traces causing the inconsistency. 
In Fig.~\ref{fig:consistency_checking}, the \textit{requires} and \textit{conflicts} traces between $r_{60}$ and $r_{59}$ are inconsistent (or contradict each other) because a requirement cannot require another requirement which it conflicts with (see the \textit{fact} in Lines 48-49 in Fig.~\ref{fig:semantics}). The inconsistent \textit{conflicts} trace is inferred using two other inferred traces. First, $r_{60}$ \textit{requires} $r_{97}$ (i.e., inferred) because \textbf{$r_{60}$} \textit{refines} $r_{11}$ which \textit{requires} $r_{97}$. Second, $r_{60}$ \textit{requires} $r_{98}$ (i.e., inferred) because $r_{60}$ \textit{requires} $r_{97}$ which contains $r_{98}$. And lastly, $r_{60}$ \textit{conflicts} with $r_{59}$ (i.e., inferred and inconsistent with the \textit{requires} trace) because $r_{60}$ \textit{requires} $r_{98}$ which \textit{conflicts} with $r_{59}$. Therefore, the manually assigned \textit{refines} trace between $r_{60}$ and $r_{11}$, \textit{requires} trace between $r_{11}$ and $r_{97}$, \textit{contains} trace between $r_{97}$ and $r_{98}$, \textit{conflicts} trace between $r_{98}$ and $r_{59}$, and \textit{requires} trace between $r_{60}$ and $r_{59}$ actually cause the inconsistency in Fig.~\ref{fig:consistency_checking}. When we, together with the Ford-Otosan engineers, analyzed all these assigned traces, we identified that the manually assigned \textit{requires} trace between $r_{60}$ and $r_{59}$ is invalid. 
We removed it to resolve the inconsistency. 

\section{Evaluation} 
\label{sec:evaluation}


Our goal was to assess, in an industrial context, the feasibility of using \T~to facilitate automated trace reasoning using user-defined trace types and semantics. For this assessment, we selected three industrial case studies which are subsystems of the ECAS system developed by different teams at Ford-Otosan~\cite{FordOtosan}. They are relatively mid-sized systems with multiple artifacts (e.g., requirement specifications, SysML models, Simulink models, test suites and C code) requiring various trace types (see Table~\ref{tab:usecase}). 

Before conducting the case studies, the Ford-Otosan engineers were given presentations illustrating the \T~steps and a tool demo. The engineers held various roles (e.g., senior software engineer and system engineer) and all had substantial experience in software development. 
For each case study, we asked the engineers to identify trace types and assisted them in specifying trace types and their semantics in \T~(the 1st and 2nd columns in Table~\ref{tab:usecase}). The artifacts in each case study had already some typeless traces (i.e., \textit{trace to}/\textit{from}) manually assigned by the engineers. We asked them to reassign those traces using the trace types they specified using \T~(the 3rd and 4th columns). 

\begin{table}[ht] 
 \vspace*{-0.8em}
\caption{Number of Trace Types, Facts, Assigned \& Inferred Traces, and Inconsistent Parts in the Case Studies} \label{tab:usecase}
 \vspace*{-1.0em}
   \begin{tabular}{c c c c  c c c}
    \toprule
       \multicolumn{1}{p{0.03\columnwidth}}{\centering } 
     & \multicolumn{1}{p{0.09\columnwidth}}{\centering Trace\\Types} 
     & \multicolumn{1}{p{0.09\columnwidth}}{\centering Facts}  
     & \multicolumn{1}{p{0.14\columnwidth}}{\centering Traced\\Elements}   
     & \multicolumn{1}{p{0.12\columnwidth}}{\centering Manual\\Traces}   
     & \multicolumn{1}{p{0.12\columnwidth}}{\centering Inferred\\Traces} 
     & \multicolumn{1}{p{0.12\columnwidth}}{\centering Inconsis.\\Parts}\\
    \midrule
    $\#1$ &7 &11 &125 &138 &502 &3\\   
    $\#2$ &11 &20 &47 &102 &145 &5\\
    $\#3$ &10 &14 &16 &21 &53 &1\\
    \bottomrule
 \end{tabular}
 \vspace*{-1.0em}
\end{table}

To evaluate the output of \T, we had semi-structured interviews with the engineers. All the inferred traces and the found inconsistencies in the case studies were confirmed by the engineers to be correct (the 5th and 6th columns). The engineers considered the automated generation of new traces and the consistency checking of traces to be highly valuable. The restricted Alloy \T~employs was sufficient to specify all the trace types and their semantics for the case studies. The engineers agreed about the useful guidance provided by \T~for specifying trace types and semantics. They stated that it was intuitive to specify trace types and semantics using \T~although more practice and training were still needed to become familiar with the tool. 

\section{Implementation \& Availability} 
\label{sec:impl}

\T~has been implemented as an Eclipse plug-in. This plug-in activates the user interfaces of \T~and provides the features \textit{specifying trace types and their semantics}, \textit{assigning traces in the artifacts using user-defined trace types}, and \textit{reasoning about traces} (i.e., \textit{deducing new traces} and \textit{checking consistency of traces}). 
We use Kodkod~\cite{Torlak:2007, KodKodPhdThesis}, an efficient SAT-based finite model finder for relational logic, to perform automated trace reasoning using the user-defined semantics. Trace types and their semantics are specified in the restricted form of Alloy, while the artifacts containing manually assigned traces are automatically transformed into Alloy specifications. Using the trace semantics and the artifacts in Alloy, we directly call KodKod API~\cite{KodkodAPI} to reason about traces.

\T~relies upon (i) a customized Eclipse editor to specify trace types and their semantics in FOL, (ii) another customized Eclipse editor to assign traces between and within the artifacts (including textual artifacts such as requirements specifications) using user-defined trace types, and (iii) \textit{alloy4graph}~\cite{Alloy4graph} and \textit{alloy4viz}~\cite{Alloy4viz}, the Alloy API packages for performing graph layout and displaying Alloy instances, to visualize the output of automated trace reasoning.

\T~is approximately 50K lines of code, excluding comments and third-party libraries.
Additional details about \T, including executable files and a screencast covering motivations, are available on the tool's website at: 
\begin{center}
\fbox{\bf
{\url{https://modelwriter.github.io/Tarski/}}}
\end{center}

\section{Conclusion} 
\label{sec:conclusion}
We presented a tool, \T, to allow the user to specify configurable trace semantics for various forms of automated trace reasoning such as inferencing and consistency checking. The key characteristics of our tool are (1) allowing the user to define new trace types and their semantics which can be later configured, (2) deducing new traces based on the traces which the user has already specified, and (3) identifying traces whose existence causes a contradiction. 
\T~has been evaluated over three industrial case studies. The evaluation shows that our tool is practical and beneficial in industrial settings to specify trace semantics for automated trace reasoning.
We plan to conduct more case studies to better evaluate the practical utility and usability of the tool.

\begin{acks}
This work is conducted within ModelWriter\cite{ModelWriter} and ASSUME\cite{Assume} projects and partially supported by the Scientific and Technological Research Council of Turkey (TUBITAK) under project \#9140014, \#9150181, and by the Luxembourg National Research Fund (FNR) (FNR/P10/03). We acknowledge networking support by European Cooperation in Science and Technology Action IC1404 "Multi-Paradigm Modelling for Cyber-Physical Systems".
\end{acks}

\newpage

\bibliographystyle{ACM-Reference-Format}
\balance
\bibliography{sigproc}


\begin{thebibliography}{00}


\ifx \showCODEN    \undefined \def \showCODEN     #1{\unskip}     \fi
\ifx \showDOI      \undefined \def \showDOI       #1{#1}\fi
\ifx \showISBNx    \undefined \def \showISBNx     #1{\unskip}     \fi
\ifx \showISBNxiii \undefined \def \showISBNxiii  #1{\unskip}     \fi
\ifx \showISSN     \undefined \def \showISSN      #1{\unskip}     \fi
\ifx \showLCCN     \undefined \def \showLCCN      #1{\unskip}     \fi
\ifx \shownote     \undefined \def \shownote      #1{#1}          \fi
\ifx \showarticletitle \undefined \def \showarticletitle #1{#1}   \fi
\ifx \showURL      \undefined \def \showURL       {\relax}        \fi
\providecommand\bibfield[2]{#2}
\providecommand\bibinfo[2]{#2}
\providecommand\natexlab[1]{#1}
\providecommand\showeprint[2][]{arXiv:#2}

\bibitem[\protect\citeauthoryear{Airbus}{Airbus}{2017}]%
        {Airbus}
\bibfield{author}{\bibinfo{person}{Airbus}.} \bibinfo{year}{2017}\natexlab{}.
\newblock \bibinfo{title}{\url{http://www.airbus.com/}}.
\newblock   (\bibinfo{year}{2017}).
\newblock


\bibitem[\protect\citeauthoryear{Aizenbud-Reshef, Paige, Rubin, Shaham-Gafni, and Kolovos}{Aizenbud-Reshef et~al\mbox{.}}{2005}]%
        {TraceabilityOperationalSemantics2005}
\bibfield{author}{\bibinfo{person}{Netta Aizenbud-Reshef}, \bibinfo{person}{Richard~F. Paige}, \bibinfo{person}{Julia Rubin}, \bibinfo{person}{Yael Shaham-Gafni}, {and} \bibinfo{person}{Dimitrios~S. Kolovos}.} \bibinfo{year}{2005}\natexlab{}.
\newblock \showarticletitle{Operational Semantics for Traceability}. In \bibinfo{booktitle}{{\em ECMDA Traceability Workshop (ECMDA-TW'05)}}. \bibinfo{pages}{8--14}.
\newblock


\bibitem[\protect\citeauthoryear{Alloy4graph}{Alloy4graph}{2017}]%
        {Alloy4graph}
\bibfield{author}{\bibinfo{person}{Alloy4graph}.} \bibinfo{year}{2017}\natexlab{}.
\newblock \bibinfo{title}{\url{http://alloy.mit.edu/alloy/documentation/alloy-api/}}.
\newblock   (\bibinfo{year}{2017}).
\newblock


\bibitem[\protect\citeauthoryear{Alloy4viz}{Alloy4viz}{2017}]%
        {Alloy4viz}
\bibfield{author}{\bibinfo{person}{Alloy4viz}.} \bibinfo{year}{2017}\natexlab{}.
\newblock \bibinfo{title}{\url{http://alloy.mit.edu/alloy/documentation/alloy-api/}}.
\newblock   (\bibinfo{year}{2017}).
\newblock


\bibitem[\protect\citeauthoryear{API}{API}{2017}]%
        {KodkodAPI}
\bibfield{author}{\bibinfo{person}{Kodkod API}.} \bibinfo{year}{2017}\natexlab{}.
\newblock \bibinfo{title}{\url{http://emina.github.io/kodkod/release/current/doc/}}.
\newblock   (\bibinfo{year}{2017}).
\newblock


\bibitem[\protect\citeauthoryear{Cleland{-}Huang, Chang, and Christensen}{Cleland{-}Huang et~al\mbox{.}}{2003}]%
        {Cleland-HuangCC03}
\bibfield{author}{\bibinfo{person}{Jane Cleland{-}Huang}, \bibinfo{person}{Carl~K. Chang}, {and} \bibinfo{person}{Mark~J. Christensen}.} \bibinfo{year}{2003}\natexlab{}.
\newblock \showarticletitle{Event-Based Traceability for Managing Evolutionary Change}.
\newblock \bibinfo{journal}{{\em {IEEE} Transactions on Software Engineering\/}} \bibinfo{volume}{29}, \bibinfo{number}{9} (\bibinfo{year}{2003}), \bibinfo{pages}{796--810}.
\newblock


\bibitem[\protect\citeauthoryear{Drivalos, Kolovos, Paige, and Fernandes}{Drivalos et~al\mbox{.}}{2008}]%
        {Drivalos2008}
\bibfield{author}{\bibinfo{person}{Nikolaos Drivalos}, \bibinfo{person}{Dimitrios~S. Kolovos}, \bibinfo{person}{Richard~F. Paige}, {and} \bibinfo{person}{Kiran~J. Fernandes}.} \bibinfo{year}{2008}\natexlab{}.
\newblock \showarticletitle{Engineering a DSL for Software Traceability}. In \bibinfo{booktitle}{{\em 1st International Conference on Software Language Engineering (SLE'08)}}. \bibinfo{pages}{151--167}.
\newblock


\bibitem[\protect\citeauthoryear{Eclipse}{Eclipse}{2017}]%
        {Eclipse}
\bibfield{author}{\bibinfo{person}{Eclipse}.} \bibinfo{year}{2017}\natexlab{}.
\newblock \bibinfo{title}{\url{https://eclipse.org}}.
\newblock   (\bibinfo{year}{2017}).
\newblock


\bibitem[\protect\citeauthoryear{Egyed}{Egyed}{2003}]%
        {Egyed03}
\bibfield{author}{\bibinfo{person}{Alexander Egyed}.} \bibinfo{year}{2003}\natexlab{}.
\newblock \showarticletitle{A Scenario-Driven Approach to Trace Dependency Analysis}.
\newblock \bibinfo{journal}{{\em {IEEE} Transactions on Software Engineering\/}} \bibinfo{volume}{29}, \bibinfo{number}{2} (\bibinfo{year}{2003}), \bibinfo{pages}{116--132}.
\newblock


\bibitem[\protect\citeauthoryear{Egyed and Gr{\"{u}}nbacher}{Egyed and Gr{\"{u}}nbacher}{2002}]%
        {EgyedG02}
\bibfield{author}{\bibinfo{person}{Alexander Egyed} {and} \bibinfo{person}{Paul Gr{\"{u}}nbacher}.} \bibinfo{year}{2002}\natexlab{}.
\newblock \showarticletitle{Automating Requirements Traceability: Beyond the Record {\&} Replay Paradigm}. In \bibinfo{booktitle}{{\em 17th {IEEE} International Conference on Automated Software Engineering {(ASE}'02)}}. \bibinfo{pages}{163--171}.
\newblock


\bibitem[\protect\citeauthoryear{Egyed and Gr{\"{u}}nbacher}{Egyed and Gr{\"{u}}nbacher}{2005}]%
        {EgyedG05}
\bibfield{author}{\bibinfo{person}{Alexander Egyed} {and} \bibinfo{person}{Paul Gr{\"{u}}nbacher}.} \bibinfo{year}{2005}\natexlab{}.
\newblock \showarticletitle{Supporting Software Understanding with Automated Requirements Traceability}.
\newblock \bibinfo{journal}{{\em International Journal of Software Engineering and Knowledge Engineering\/}} \bibinfo{volume}{15}, \bibinfo{number}{5} (\bibinfo{year}{2005}), \bibinfo{pages}{783--810}.
\newblock


\bibitem[\protect\citeauthoryear{Erata, Challenger, Tekinerdogan, Monceaux, T\"{u}z\"{u}n, and Kardas}{Erata et~al\mbox{.}}{2017}]%
        {ferhat2017SAC:PL}
\bibfield{author}{\bibinfo{person}{Ferhat Erata}, \bibinfo{person}{Moharram Challenger}, \bibinfo{person}{Bedir Tekinerdogan}, \bibinfo{person}{Anne Monceaux}, \bibinfo{person}{Eray T\"{u}z\"{u}n}, {and} \bibinfo{person}{Geylani Kardas}.} \bibinfo{year}{2017}\natexlab{}.
\newblock \showarticletitle{Tarski: A Platform for Automated Analysis of Dynamically Configurable Traceability Semantics}. In \bibinfo{booktitle}{{\em Proceedings of the Symposium on Applied Computing}} {\em (\bibinfo{series}{SAC '17})}. \bibinfo{publisher}{ACM}, \bibinfo{address}{New York, NY, USA}, \bibinfo{pages}{1607--1614}.
\newblock
\showISBNx{978-1-4503-4486-9}
\showDOI{%
\url{https://doi.org/10.1145/3019612.3019747}}


\bibitem[\protect\citeauthoryear{for European~Advancement)}{for European~Advancement)}{2014}]%
        {ModelWriter}
\bibfield{author}{\bibinfo{person}{{ITEA} (Information~Technology for European~Advancement)}.} \bibinfo{year}{2014}\natexlab{}.
\newblock \bibinfo{title}{{ModelWriter}: Text \& Model Synchronized Document Engineering Platform}.
\newblock \bibinfo{howpublished}{\url{https://itea3.org/project/modelwriter.html}}.   (\bibinfo{date}{Sep} \bibinfo{year}{2014}).
\newblock


\bibitem[\protect\citeauthoryear{for European~Advancement)}{for European~Advancement)}{2015}]%
        {Assume}
\bibfield{author}{\bibinfo{person}{{ITEA} (Information~Technology for European~Advancement)}.} \bibinfo{year}{2015}\natexlab{}.
\newblock \bibinfo{title}{{ASSUME}: Affordable Safe \& Secure Mobility Evolution}.
\newblock \bibinfo{howpublished}{\url{https://itea3.org/project/assume.html}}.   (\bibinfo{date}{Sep} \bibinfo{year}{2015}).
\newblock


\bibitem[\protect\citeauthoryear{Ford-Otosan}{Ford-Otosan}{2017}]%
        {FordOtosan}
\bibfield{author}{\bibinfo{person}{Ford-Otosan}.} \bibinfo{year}{2017}\natexlab{}.
\newblock \bibinfo{title}{\url{http://www.fordotosan.com.tr}}.
\newblock   (\bibinfo{year}{2017}).
\newblock


\bibitem[\protect\citeauthoryear{Goknil, Kurtev, and Berg}{Goknil et~al\mbox{.}}{2014}]%
        {FormalTraceSemantics2014}
\bibfield{author}{\bibinfo{person}{Arda Goknil}, \bibinfo{person}{Ivan Kurtev}, {and} \bibinfo{person}{Klaas Van~Den Berg}.} \bibinfo{year}{2014}\natexlab{}.
\newblock \showarticletitle{Generation and Validation of Traces between Requirements and Architecture based on Formal Trace Semantics}.
\newblock \bibinfo{journal}{{\em Journal of Systems and Software\/}}  \bibinfo{volume}{88} (\bibinfo{year}{2014}), \bibinfo{pages}{112--137}.
\newblock


\bibitem[\protect\citeauthoryear{Goknil, Kurtev, and Millo}{Goknil et~al\mbox{.}}{2013}]%
        {GoknilKM13}
\bibfield{author}{\bibinfo{person}{Arda Goknil}, \bibinfo{person}{Ivan Kurtev}, {and} \bibinfo{person}{Jean{-}Vivien Millo}.} \bibinfo{year}{2013}\natexlab{}.
\newblock \showarticletitle{A Metamodeling Approach for Reasoning on Multiple Requirements Models}. In \bibinfo{booktitle}{{\em 17th {IEEE} International Enterprise Distributed Object Computing Conference (EDOC'13)}}. \bibinfo{pages}{159--166}.
\newblock


\bibitem[\protect\citeauthoryear{G{\"o}knil, Kurtev, and van~den Berg}{G{\"o}knil et~al\mbox{.}}{2008}]%
        {goknil2008change}
\bibfield{author}{\bibinfo{person}{Arda G{\"o}knil}, \bibinfo{person}{Ivan Kurtev}, {and} \bibinfo{person}{Klaas van~den Berg}.} \bibinfo{year}{2008}\natexlab{}.
\newblock \showarticletitle{Change Impact Analysis based on Formalization of Trace Relations for Requirements}. In \bibinfo{booktitle}{{\em the {ECMDA} Traceability Workshop (ECMDA-TW'08)}}. \bibinfo{pages}{59--75}.
\newblock


\bibitem[\protect\citeauthoryear{Goknil, Kurtev, and van~den Berg}{Goknil et~al\mbox{.}}{2008}]%
        {goknil2008metamodeling}
\bibfield{author}{\bibinfo{person}{Arda Goknil}, \bibinfo{person}{Ivan Kurtev}, {and} \bibinfo{person}{Klaas van~den Berg}.} \bibinfo{year}{2008}\natexlab{}.
\newblock \showarticletitle{A Metamodeling Approach for Reasoning about Requirements}. In \bibinfo{booktitle}{{\em European Conference on Model Driven Architecture-Foundations and Applications (ECMDA-FA'08)}}. \bibinfo{pages}{310--325}.
\newblock


\bibitem[\protect\citeauthoryear{Goknil, Kurtev, and van~den Berg}{Goknil et~al\mbox{.}}{2010}]%
        {GoknilKB10}
\bibfield{author}{\bibinfo{person}{Arda Goknil}, \bibinfo{person}{Ivan Kurtev}, {and} \bibinfo{person}{Klaas van~den Berg}.} \bibinfo{year}{2010}\natexlab{}.
\newblock \showarticletitle{Tool Support for Generation and Validation of Traces between Requirements and Architecture}. In \bibinfo{booktitle}{{\em the 6th {ECMFA} Traceability Workshop (ECMFA-TW'10)}}. \bibinfo{pages}{39--46}.
\newblock


\bibitem[\protect\citeauthoryear{Goknil, Kurtev, van~den Berg, and Spijkerman}{Goknil et~al\mbox{.}}{2014}]%
        {ChangeImpact2014}
\bibfield{author}{\bibinfo{person}{Arda Goknil}, \bibinfo{person}{Ivan Kurtev}, \bibinfo{person}{Klaas van~den Berg}, {and} \bibinfo{person}{Wietze Spijkerman}.} \bibinfo{year}{2014}\natexlab{}.
\newblock \showarticletitle{Change Impact Analysis for Requirements: A Metamodeling Approach}.
\newblock \bibinfo{journal}{{\em Information and Software Technology\/}} \bibinfo{volume}{56}, \bibinfo{number}{8} (\bibinfo{year}{2014}), \bibinfo{pages}{950 -- 972}.
\newblock


\bibitem[\protect\citeauthoryear{Goknil, Kurtev, van~den Berg, and Veldhuis}{Goknil et~al\mbox{.}}{2011}]%
        {SemanticTrace2011}
\bibfield{author}{\bibinfo{person}{Arda Goknil}, \bibinfo{person}{Ivan Kurtev}, \bibinfo{person}{Klaas van~den Berg}, {and} \bibinfo{person}{Jan-Willem Veldhuis}.} \bibinfo{year}{2011}\natexlab{}.
\newblock \showarticletitle{Semantics of Trace Relations in Requirements Models for Consistency Checking and Inferencing}.
\newblock \bibinfo{journal}{{\em Software and System Modeling\/}} \bibinfo{volume}{10}, \bibinfo{number}{1} (\bibinfo{year}{2011}), \bibinfo{pages}{31--54}.
\newblock


\bibitem[\protect\citeauthoryear{Gyawali, Shimorina, Gardent, Cruz-Lara, and Mahfoudh}{Gyawali et~al\mbox{.}}{2017}]%
        {Gyawali2017}
\bibfield{author}{\bibinfo{person}{Bikash Gyawali}, \bibinfo{person}{Anastasia Shimorina}, \bibinfo{person}{Claire Gardent}, \bibinfo{person}{Samuel Cruz-Lara}, {and} \bibinfo{person}{Mariem Mahfoudh}.} \bibinfo{year}{2017}\natexlab{}.
\newblock \showarticletitle{Mapping Natural Language to Description Logic}. In \bibinfo{booktitle}{{\em 14th European Semantic Web Conference (ESWC'17)}}. \bibinfo{pages}{273--288}.
\newblock


\bibitem[\protect\citeauthoryear{Havelsan}{Havelsan}{2017}]%
        {Havelsan}
\bibfield{author}{\bibinfo{person}{Havelsan}.} \bibinfo{year}{2017}\natexlab{}.
\newblock \bibinfo{title}{\url{http://www.havelsan.com.tr}}.
\newblock   (\bibinfo{year}{2017}).
\newblock


\bibitem[\protect\citeauthoryear{ISO}{ISO}{2017}]%
        {ISO26262}
\bibfield{author}{\bibinfo{person}{ISO}.} \bibinfo{year}{2017}\natexlab{}.
\newblock \bibinfo{title}{{ISO-26262}: Road vehicles -- Functional safety}.
\newblock   (\bibinfo{year}{2017}).
\newblock


\bibitem[\protect\citeauthoryear{Jackson}{Jackson}{2012}]%
        {AlloyBook}
\bibfield{author}{\bibinfo{person}{Daniel Jackson}.} \bibinfo{year}{2012}\natexlab{}.
\newblock \bibinfo{booktitle}{{\em Software Abstractions: Logic, Language, and Analysis}}.
\newblock \bibinfo{publisher}{MIT press}.
\newblock


\bibitem[\protect\citeauthoryear{Kolovos, Paige, and Polack}{Kolovos et~al\mbox{.}}{2008}]%
        {DetectRepairConsistency2008}
\bibfield{author}{\bibinfo{person}{Dimitrios~S. Kolovos}, \bibinfo{person}{Richard~F. Paige}, {and} \bibinfo{person}{Fiona Polack}.} \bibinfo{year}{2008}\natexlab{}.
\newblock \showarticletitle{Detecting and Repairing Inconsistencies across Heterogeneous Models}. In \bibinfo{booktitle}{{\em 1st International Conference on Software Testing, Verification, and Validation}}. \bibinfo{pages}{356--364}.
\newblock
\showISSN{2159-4848}


\bibitem[\protect\citeauthoryear{Lamb, Jirapanthong, and Zisman}{Lamb et~al\mbox{.}}{2011}]%
        {lamb2011}
\bibfield{author}{\bibinfo{person}{Luis~C. Lamb}, \bibinfo{person}{Waraporn Jirapanthong}, {and} \bibinfo{person}{Andrea Zisman}.} \bibinfo{year}{2011}\natexlab{}.
\newblock \showarticletitle{Formalizing Traceability Relations for Product Lines}. In \bibinfo{booktitle}{{\em the 6th International Workshop on Traceability in Emerging Forms of Software Engineering (TEFSE'11)}}. \bibinfo{pages}{42--45}.
\newblock


\bibitem[\protect\citeauthoryear{Ramesh and Jarke}{Ramesh and Jarke}{2001}]%
        {Ramesh:2001aa}
\bibfield{author}{\bibinfo{person}{Balasubramaniam Ramesh} {and} \bibinfo{person}{Matthias Jarke}.} \bibinfo{year}{2001}\natexlab{}.
\newblock \showarticletitle{Toward Reference Models for Requirements Traceability}.
\newblock \bibinfo{journal}{{\em IEEE Transactions on Software Engineering\/}} \bibinfo{volume}{27}, \bibinfo{number}{1} (\bibinfo{year}{2001}), \bibinfo{pages}{58--93}.
\newblock


\bibitem[\protect\citeauthoryear{RTCA and EUROCAE}{RTCA and EUROCAE}{2017}]%
        {DO178C}
\bibfield{author}{\bibinfo{person}{RTCA} {and} \bibinfo{person}{EUROCAE}.} \bibinfo{year}{2017}\natexlab{}.
\newblock \bibinfo{title}{{DO-178C}: Software Considerations in Airborne Systems and Equipment Certification}.
\newblock   (\bibinfo{year}{2017}).
\newblock


\bibitem[\protect\citeauthoryear{Sabetzadeh, Nejati, Liaskos, Easterbrook, and Chechik}{Sabetzadeh et~al\mbox{.}}{2007}]%
        {ConsistencyCheckingModelMerging}
\bibfield{author}{\bibinfo{person}{Mehrdad Sabetzadeh}, \bibinfo{person}{Shiva Nejati}, \bibinfo{person}{Sotirios Liaskos}, \bibinfo{person}{Steve Easterbrook}, {and} \bibinfo{person}{Marsha Chechik}.} \bibinfo{year}{2007}\natexlab{}.
\newblock \showarticletitle{Consistency Checking of Conceptual Models via Model Merging}. In \bibinfo{booktitle}{{\em 15th IEEE International Requirements Engineering Conference (RE'07)}}. \bibinfo{pages}{221--230}.
\newblock


\bibitem[\protect\citeauthoryear{Society, Bourque, and Fairley}{Society et~al\mbox{.}}{2014}]%
        {SWEBOK:IEEEComputerSociety:2014}
\bibfield{author}{\bibinfo{person}{IEEE~Computer Society}, \bibinfo{person}{Pierre Bourque}, {and} \bibinfo{person}{Richard~E. Fairley}.} \bibinfo{year}{2014}\natexlab{}.
\newblock \bibinfo{booktitle}{{\em Guide to the Software Engineering Body of Knowledge (SWEBOK(R)): Version 3.0\/} (\bibinfo{edition}{3rd} ed.)}.
\newblock \bibinfo{publisher}{IEEE Computer Society Press}, \bibinfo{address}{Los Alamitos, CA, USA}.
\newblock
\showISBNx{0769551661, 9780769551661}


\bibitem[\protect\citeauthoryear{Tarski}{Tarski}{1941}]%
        {tarski1941calculus}
\bibfield{author}{\bibinfo{person}{Alfred Tarski}.} \bibinfo{year}{1941}\natexlab{}.
\newblock \showarticletitle{On the Calculus of Relations}.
\newblock \bibinfo{journal}{{\em The Journal of Symbolic Logic\/}} \bibinfo{volume}{6}, \bibinfo{number}{03} (\bibinfo{year}{1941}), \bibinfo{pages}{73--89}.
\newblock


\bibitem[\protect\citeauthoryear{ten Hove, Goknil, Kurtev, van~den Berg, and de~Goede}{ten Hove et~al\mbox{.}}{2009}]%
        {hove2009change}
\bibfield{author}{\bibinfo{person}{David ten Hove}, \bibinfo{person}{Arda Goknil}, \bibinfo{person}{Ivan Kurtev}, \bibinfo{person}{Klaas van~den Berg}, {and} \bibinfo{person}{Koos de Goede}.} \bibinfo{year}{2009}\natexlab{}.
\newblock \showarticletitle{Change Impact Analysis for {SysML} Requirements Models based on Semantics of Trace Relations}. In \bibinfo{booktitle}{{\em the {ECMDA} Traceability Workshop (ECMDA-TW'09)}}. \bibinfo{pages}{17--28}.
\newblock


\bibitem[\protect\citeauthoryear{Torlak}{Torlak}{2008}]%
        {KodKodPhdThesis}
\bibfield{author}{\bibinfo{person}{Emina Torlak}.} \bibinfo{year}{2008}\natexlab{}.
\newblock {\em \bibinfo{title}{A Constraint Solver for Software Engineering: Finding Models and Cores of Large Relational Specifications}}.
\newblock \bibinfo{thesistype}{Ph.D. Dissertation}. \bibinfo{school}{Massachusetts Institute of Technology}.
\newblock


\bibitem[\protect\citeauthoryear{Torlak and Jackson}{Torlak and Jackson}{2007}]%
        {Torlak:2007}
\bibfield{author}{\bibinfo{person}{Emina Torlak} {and} \bibinfo{person}{Daniel Jackson}.} \bibinfo{year}{2007}\natexlab{}.
\newblock \showarticletitle{Kodkod: A Relational Model Finder}. In \bibinfo{booktitle}{{\em the 13th International Conference on Tools and Algorithms for the Construction and Analysis of Systems (TACAS'07)}}. \bibinfo{pages}{632--647}.
\newblock


\end{thebibliography}
\newpage
\section*{Appendices}

We provide three appendices for the paper. In Appendix \ref{apx:availability}, we provide source code, tool and data availability details. In Appendix \ref{apx:demonstration}, we explain a walk through of the actual presentation in details. Finally in Appendix \ref{apx:axiomatization}, we present a full axiomatization of the industrial use case in predicate calculus style for interested readers.

\appendix

\section{Availability \& Open Source License} \label{apx:availability} 

\noindent\textbf{Source Codes, Screencast and Datasets}. The source codes files and datasets of \T~are publicly available for download and use at the project website. A screencast and the installation steps for \T~are also available at the same website and can be found at: 

\begin{center}
\fbox{\bf{\url{https://modelwriter.github.io/Tarski/}}}
\end{center}

\noindent\T~ is being developed under Work Package 3 within ModelWriter project, labeled by the European Union's EUREKA Cluster programme ITEA (Information Technology for European Advancement). Further details about the project can be found at: 

\begin{center}
\fbox{\bf{\url{https://itea3.org/project/modelwriter.html}}}
\end{center}

\noindent\textbf{Open Source License}. Tarski is distributed with an open source software license, namely \textit{Eclipse Public License v1}. This commercially friendly copyleft license provides the ability to commercially license binaries; a modern royalty-free patent license grant; and the ability for linked works to use other licenses, including commercial ones.

\begin{center}
\fbox{\bf{\url{https://github.com/ModelWriter/Tarski/blob/master/LICENSE}}}
\end{center}

\section{Tool Demonstration Plan} \label{apx:demonstration} 
There will be four parts to our presentation: (1) motivation and industrial use cases, (2) overview of the solution and tool architecture, (3) demonstration walktrough, and (4) evaluation.  Parts 1, 2 and 4 are presented using slides while Part 3 is presented as a demo using the industrial use case scenario described in Section \ref{sec:overview} and detailed in Appendix \ref{apx:axiomatization}. To present these parts, we use a combination of slides, animations, and a live demo. In the following subsections, we provide further details about our presentation plan. 

A 25-minute slot has been assumed for the presentation. The estimated duration for the different parts of the presentation suggested below will be adjusted proportionally if the allocated time slot at the conference is different from the above.

\subsection{Motivation \& Challenges} \label{Motivation}
\textit{Estimated Duration: 4 - 5 minutes.}\\
\textit{Delivery: 3 to 4 slides.}

\noindent\textbf{Motivation.} We will emphasize the importance of traceability by introducing "\textit{DO-178C Software Considerations in Airborne Systems and Equipment Certification}"~\cite{DO178C} from aviation industry and  "\textit{ISO-26262 Road vehicles - Functional safety}"~\cite{ISO26262} from automotive Industry. 

\noindent\textbf{Industrial Use Cases} We will briefly describe the challenges of \textit{Traceability Analysis Activities} faced in industry by introducing industrial use cases from Airbus~\cite{Airbus}, Ford-Otosan~\cite{FordOtosan} and Havelsan~\cite{Havelsan}. We will explain the importance of \textit{semantically meaningful traceability} and \textit{traceability configuration} in industry.

\subsection{Tool Overview} \label{Overview}
\textit{Estimated Duration: 1 - 2 minutes.}\\
\textit{Delivery: 2 to 3 slides.}

\noindent\textbf{Overview of the Solution.} We will explain the approach and the user workflow of \T~ by following the steps as shown in Fig.~\ref{fig:tool_overview}.

\noindent\textbf{\T~Features.} We will briefly explain tool features such as \textit{\nameref{subsubsec:inference}} and \textit{\nameref{subsubsec:consistency}} using animated slides by giving concrete examples from the industrial use case \E~ system presented in the paper. Apart from those features, \T~provides two more analysis functions, which are \textit{Type Approximation for Trace-Locations} and \textit{Reasoning about a Specific Trace-Location}.

\subsection{Walk-trough of the Tool Demonstration} \label{Demonstration}
\textit{Estimated Duration: 9 - 11 minutes.}\\
\textit{Delivery: 3 to 4 slides together with a live demo.}

\noindent\textbf{Tool Demonstration.} In this section, first, we will describe the traceability domain model of the industrial use case, namely \case, which is illustrated in Fig. \ref{fig:traceability_domain_model} and formalized in Fig. \ref{fig:specification_types} and Fig. \ref{fig:semantics} using \T. We will perform a live demonstration taking the following steps: 
\begin{enumerate}[leftmargin=*] 

\item {

\textbf{Introduction of the Eclipse workspace.} A workspace in Eclipse contains a collection of resources, i.e. projects, folders and files. In Fig. \ref{step:workspace}, in the user workspace there exist several projects, one of them contains a requirement specification file, a source code file and an architectural model. 

\begin{figure}[H] 
\centering
\includegraphics[width=\columnwidth]{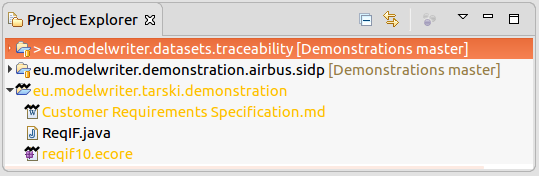}
\caption{User's Eclipse Workspace} \label{step:workspace}
\end{figure}

}

\item {

\textbf{Project-based configuration of \textit{trace-location} types.} In Fig. \ref{step:configuration_types}, user declares a type hierarchy by creating trace types, which are explained in Section \ref{subsec:specification}. The type hierarchy for trace locations in Fig. \ref{step:configuration_types} consists of three sorts, namely \type{Requirement}, \type{Implementation} and \type{Specificaton}. \type{HighLevelReq} and \type{LowLevelReq} is sub-types of \type{Requirement}, whereas \type{Model} and \type{Code} are sub-types of \type{Implementation}.

\begin{figure}[H]
\centering
\includegraphics[width=\columnwidth]{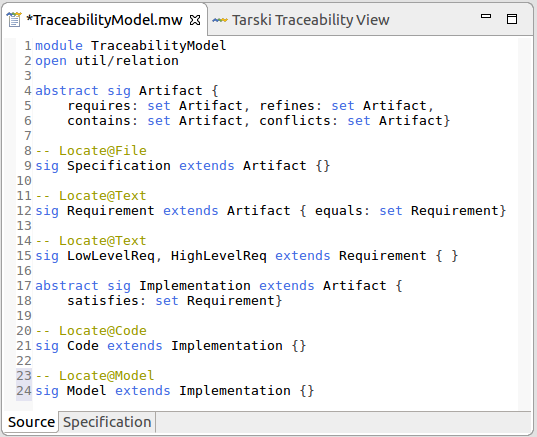}
\caption{Some Example Trace Types in \T}
\label{step:configuration_types}
\end{figure}

\noindent Each \T~ specification configures the Eclipse IDE to trace artifacts in a specific Eclipse workspace. In Fig. \ref{step:TypeHierarchy}, the type hierarchy of the configuration file is visualized using the editor. The user can drag and drop selected \textit{text} and \textit{code} fragments onto types to create a trace-locations. Since in Alloy formalism, $abstract$ signatures have no elements except those belonging to its extensions, the system does not allow the user to create a trace-location with an abstract type such as $Implementation$ or $Artefact$.

\begin{figure}[H]
\centering
\includegraphics[width=\columnwidth]{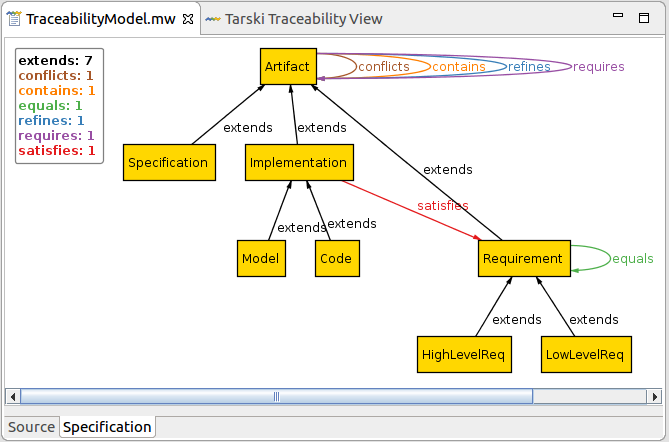}
\caption{Trace Type Hierarchy of \E}
\label{step:TypeHierarchy}
\end{figure}

}

\item {

\textbf{Project-based configuration of the trace semantics.} In Fig. \ref{step:configuration_semantics}, the user defines several consistency rules such as \textit{injectivity} of \type{contains} trace, and \textit{reflexivity} and \textit{anti-symmetry} of \type{refines}, \type{contains} and \type{requires}. The user also selects \type{conflicts}, \type{satisfies} and \type{requires} traces for inferring new trace relations.
    
\begin{figure}[H]
\centering
\includegraphics[width=\columnwidth]{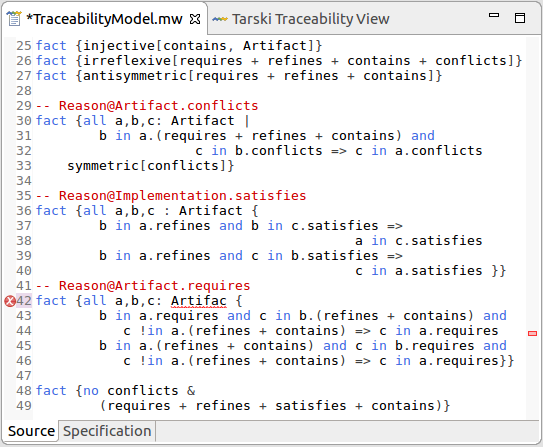}
\caption{An Example Trace Semantics in \T}
\label{step:configuration_semantics}
\end{figure}
}

\item\label{step:4} {

\textbf{Uploading configuration file.} User upload the configuration file to \T~ plug-in using the menu item as shown in Fig.~\ref{step:uploading}.

\begin{figure}[H]
\centering
\includegraphics[width=\columnwidth]{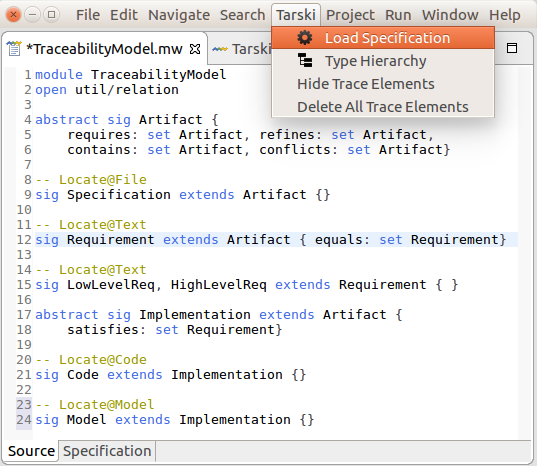}
\caption{Uploading the configuration file}
\label{step:uploading}
\end{figure}
}

\item {

\textbf{Tracing different parts of various artifacts.} In \T, each \textit{trace-location} and \textit{trace-link} subject to formal analysis must be annotated with a type from the hierarchy obtained from the $signature$ and the $field$ declarations on the specification (e.g. Fig.~\ref{step:configuration_types}). In the user's workspace, for instance, user traces text fragments of a requirement specification in Fig. \ref{step:TextLocation}, several model elements of an architecture document in Fig. \ref{step:XMILocation}, and several language constructs of a source code file in Fig. \ref{step:JavaLocation}.

\begin{figure}[h] 
\centering
\includegraphics[width=\columnwidth]{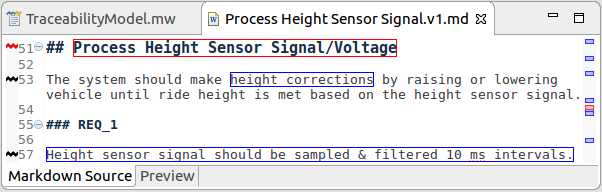}
\caption{Traced locations in the requirements document} \label{step:TextLocation}
\end{figure}

\begin{figure}[h] 
\centering
\includegraphics[width=\columnwidth]{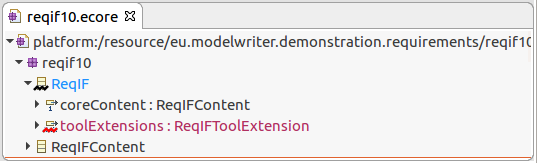}
\caption{Traced locations in the architecture model} \label{step:XMILocation}
\end{figure}

\begin{figure}[h] 
\centering
\includegraphics[width=\columnwidth]{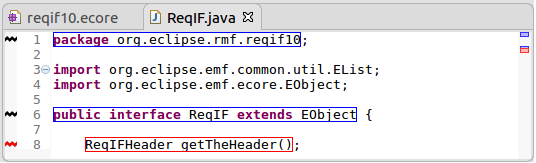}
\caption{Traced locations in the source codes} \label{step:JavaLocation}
\end{figure}

}

\item {

\textbf{Creating trace-relation with type.} First, the user selects a trace-location which constitutes the domain of the intended relation using the context-menu. In Fig. \ref{step:select_location}, (s)he selects a \type{code} typed trace-location.

\begin{figure}[h] 
\centering
\includegraphics[width=\columnwidth]{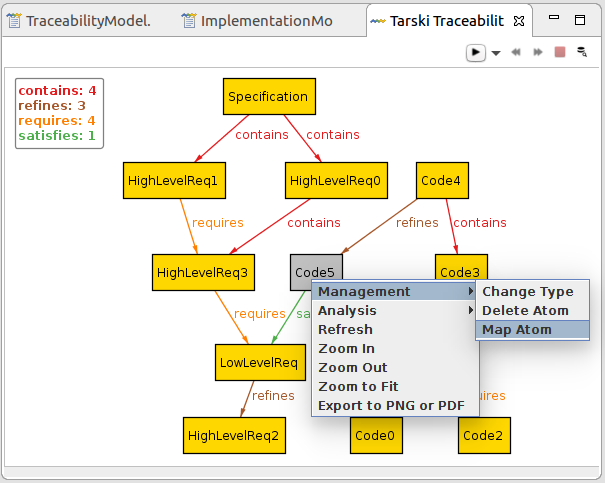}
\caption{Select a trace-location for the domain} \label{step:select_location}
\end{figure}

A wizard pops up to list legitimate trace types as shown in Fig. \ref{step:select_location}. \T~ is able to resolve subtype polimorphism and suggests suitable trace-types. In this case, the user selects the type of \type{satisfies}. 

\begin{figure}[h] 
\centering
\includegraphics[width=\columnwidth]{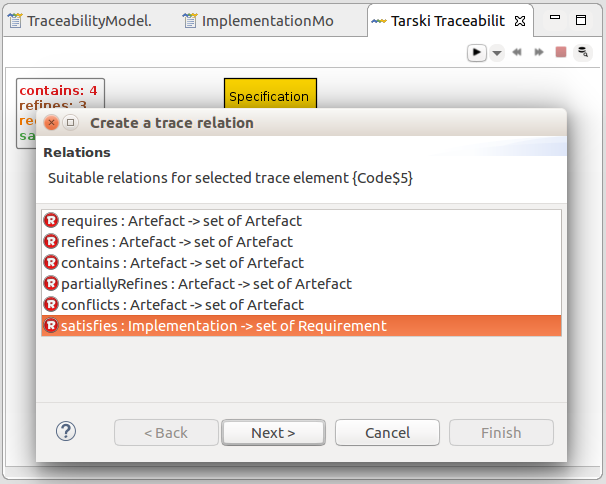}
\caption{Select a trace-type} \label{step:select_type}
\end{figure}

In the next screen of the wizard, system proposes available trace-locations as the range of the relation considering the trace type hierarchy that user specified previously through the configuration file. He selects \textit{Code\$0} from the requirement document as shown in Fig. \ref{step:select_range}. 

\begin{figure}[h] 
\centering
\includegraphics[width=\columnwidth]{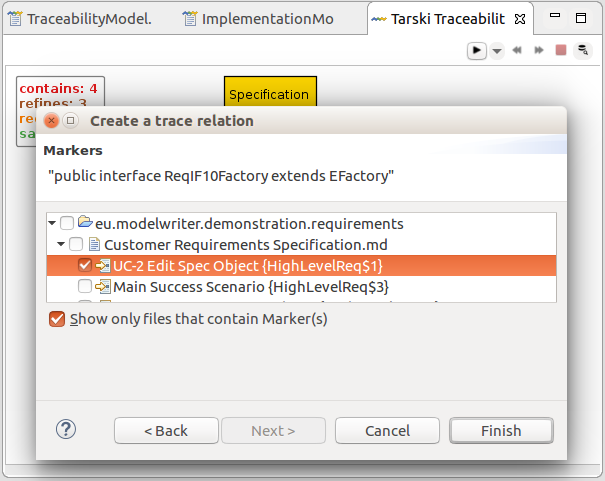}
\caption{Select a trace-location for the range} \label{step:select_range}
\end{figure}

}

\item {

\textbf{Traceability management.} The traceability information is adapted to a first-order relational model by the user's type annotations to \textit{location}s (unary relations) and \textit{traces} (binary. ternary and n-ary relations). Each user function has a counter-part API method in order to create automatically those trace-elements especially in model-based development. Furthermore, Tarski platform provides functions such as \textit{create}, \textit{delete}, \textit{update} and \textit{change type} of the relations with respect to type hierarchy and multiplicity constraints to enable users to elaborate further on the formal instance. The user can manage the \textit{locations} and \textit{traces} using \T~ Traceability Visualization View as shown in Fig. \ref{step:trace_management}.

\begin{figure}[h] 
\centering
\includegraphics[width=\columnwidth]{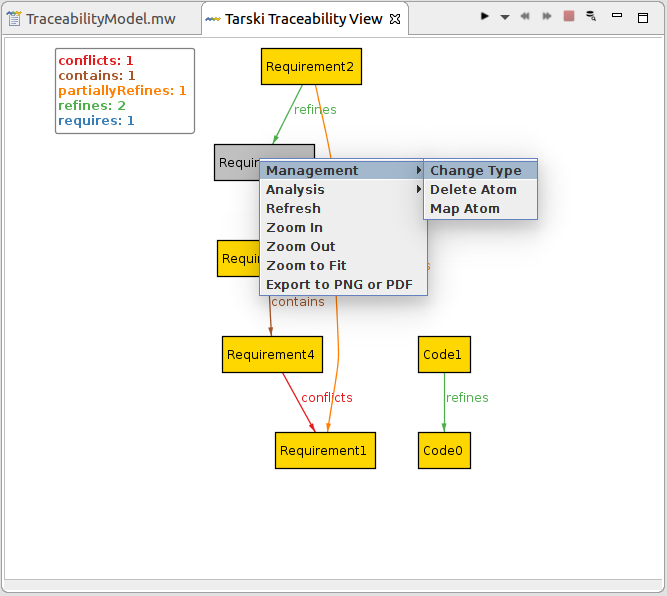}
\caption{Traces on the current snapshot} \label{step:trace_management}
\end{figure}

The user is also able to change the type of a location as shown in Fig.~\ref{step:TypeAnnotation}.

\begin{figure}[h] 
\centering
\includegraphics[width=\columnwidth]{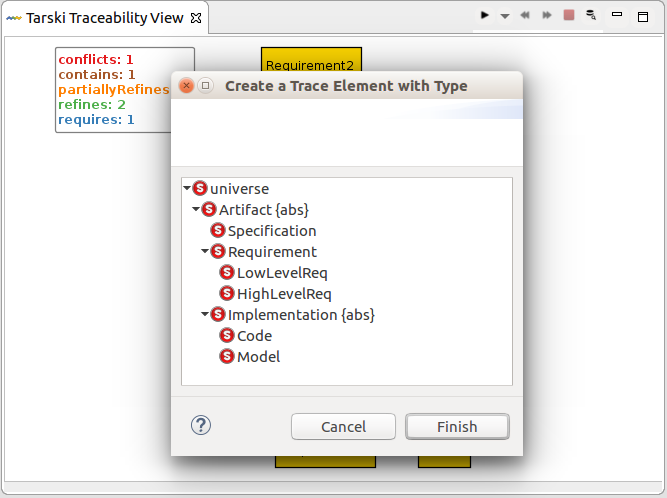}
\caption{Annotate type to a traced \textit{location}} \label{step:TypeAnnotation}
\end{figure}

}

\item  {

\textbf{Consistency checking.} The user can check the consistency on the existing traces and detect an inconsistency as explained in Section \ref{subsubsec:consistency}. The user is informed if such an inconsistency occurs (Fig. \ref{step:inconsistency}).

\begin{figure}[h] 
\centering
\includegraphics[width=\columnwidth]{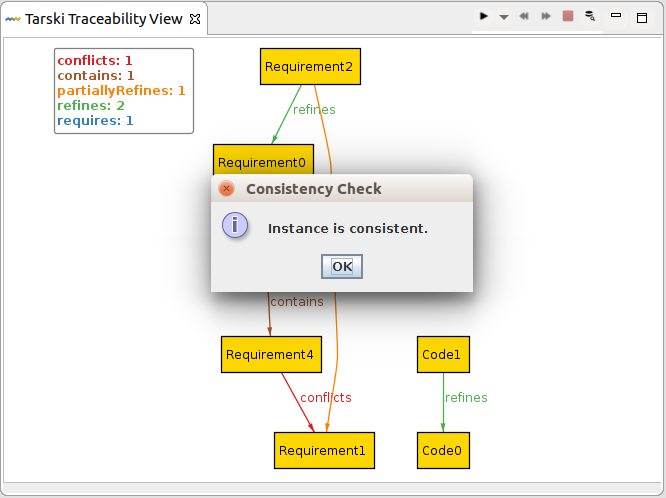}
\caption{Traces on the current snapshot} \label{step:inconsistency}
\end{figure}

}

\item{ 

\textbf{Inferring new traces.} The details of inferring new traces are explained in Section \ref{subsubsec:inference}. If the user performs reasoning operations about traces, the result is reported back to the user by dashed traces as shown in Fig. \ref{step:inference}. If there exists different solutions, the user can traverse them back and forth. He can also accept the inferred traces, and perform another analysis operation including inferred traces. 

\begin{figure}[h]
\includegraphics[width=\columnwidth]{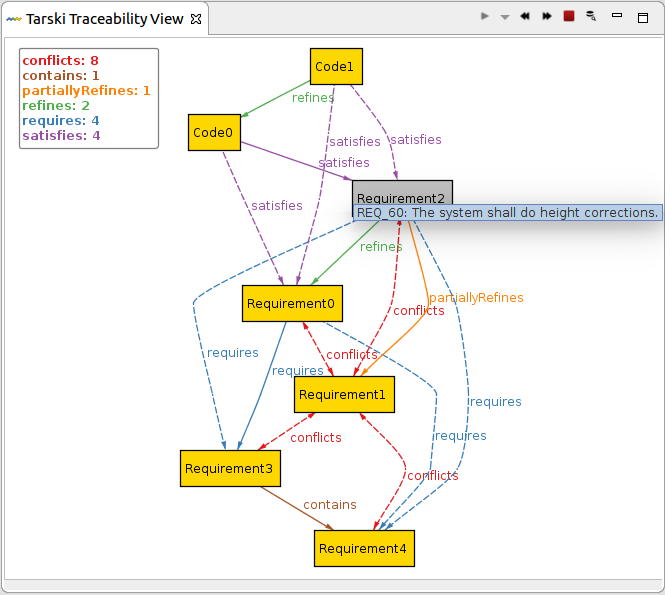}
\caption{Inferred Relations based on the current snapshot} \label{step:inference}
\end{figure}

}

\item {

\textbf{Reconfiguring the trace semantics.} Using the current traces the user is able to change the trace semantics and submit the new specification to the system unless (s)he changes trace types. In this way, (s)he can generate new traces adopting different semantics based on his/her project's changing needs. The user then continues the analysis process with Step \ref{step:4}.

}

\end{enumerate}

\subsection{Evaluation and Lessons Learned} \label{Evaluation}

\textit{Estimated Duration: 1 - 2 minutes.}\\
\textit{Delivery: 2 slides.}


\noindent We conclude with a summary that presents the evaluation results and the lessons learned.

\section{Axiomatization of the Case Study} \label{apx:axiomatization}

In this section, we axiomatize trace semantics of the case study using \textit{First-order Predicate Logic} with the signature:
\begin{gather*}
    \Sigma_T : \{ =, \in \} \cup \Sigma^1_T \cup \Sigma^2_T \\
    \Sigma^1_T : \{ Artifact, Requirement, Implementation \}\\
    \Sigma^2_T : \{ requires, refines, contains, equals, conflicts, satisfies \}
\end{gather*}
$\Sigma^1_T$ is the set of unary predicate symbols and $\Sigma^2_T$ is the set of binary predicate symbols. For simplicity, we assume that the universe only consists of the type, \textit{Artifact} which is partitioned into disjoint subsets of \textit{Requirement} and \textit{Implementation}. From now on, $A$ represents the set of \textit{Artifacts}.

\subsection{Informal Definitions of Trace-types} \label{apx:sub:definitions} 

In the following list, we informally give the meaning of the trace-types:

\begin{enumerate} [wide, labelwidth=!, labelindent=0pt]

\item \textbf{\textit{requires}} \textit{Artifact} $A_1$ \textit{requires} \textit{Artifact} $A_2$ if $A_1$ is fulfilled only when $A_2$ is fulfilled. The required artifact can be seen as a pre-condition for the requiring artifact.

\item \textbf{\textit{contains.}} \textit{Artifact} $A_1$ \textit{contains} \textit{Artifacts} $A_2 \dots A_n$ if $A_2 \dots A_n$ are parts of the whole $A_1$ (part-whole hierarchy). 

\item \textbf{\textit{refines.}} \textit{Artifact} $A_1$ \textit{refines} another \textit{Artifact} $A_2$ if $A_1$ is derived from $A_2$ by adding more details to its properties. The refined artifact can be seen as an abstraction of the detailed artifacts.

\item \textbf{\textit{conflicts.}} \textit{Artifact} $A_1$ \textit{conflicts} with \textit{Artifact} $A_2$ if the fulfillment of $A_1$ excludes the fulfillment of $A_2$ and vice versa.

\item \textbf{\textit{equals.}} \textit{Artifact} $A_1$ \textit{equals} to \textit{Artifact} $A_2$ if $A_1$ states exactly the same properties with their constraints with $A_2$ and vice versa.

\item \textbf{\textit{satisfies.}} \textit{Implementation} $A_1$ \textit{satisfies} \textit{Requirement} $A_2$ if $A_1$ implements all the properties stated by $A_2$ . 

\end{enumerate}

\subsection{Formal Semantics of Trace-types} \label{apx:sub:formalization} 

In the following several axiom schemas are listed to formalize Traceability Theory, that is used in the \E~ case study.

\begin{enumerate}[leftmargin=*]

\item Reasoning about \textit{equals} relation, the pattern in Fig. \ref{fig:reasoning_equality} is used.

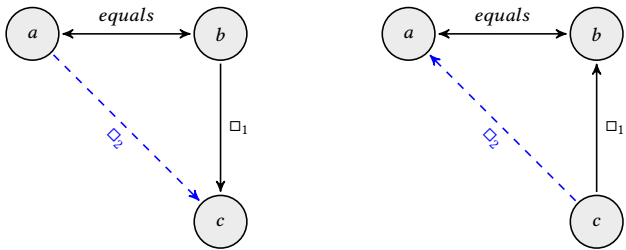
\begin{figure}[H]
\centering
      \begin{tikzpicture}[modal,node distance=2.5cm,world/.append style={minimum size=0.7cm}]

        \node (A) [world] {$a$};
        \node (B) [world, right of=A] {$b$};
        \node (C) [world, below of=B] {$c$};
        \node (D) [world, right of=B] {$a$};
        \node (E) [world, right of=D] {$b$};
        \node (F) [world, below of=E] {$c$};

        \path [<->]
            (A) edge node {$equals$} (B)
            (D) edge node {$equals$} (E);

        \path [->]
            (B) edge node {$\square_1$} (C)
            (F) edge node[right] {$\square_1$} (E);

        \path [->] [blue, dashed, >=stealth']
            (A) edge node[sloped, anchor=north] {$\square_2$} (C)
            (F) edge node[sloped, anchor=north] {$\square_2$} (D);

      \end{tikzpicture}
\caption{Patterns to reason about "\textit{equality}"}
\label{fig:reasoning_equality}
\end{figure}

\begin{gather}\label{axioms_equals} 
    \vdash \forall a, b, c \in A \mid (a,b) \in \equals \wedge\, (b,c) \in \square \to (a,c) \in \square \\
    \vdash \forall a, b, c \in A \mid (a,b) \in \equals \wedge\, (c,b) \in \square \to (c,a) \in \square \\
    \vdash \forall a \in A \mid (a,a) \in \equals \\
    \notag\mbox{where \:}  \square \in \{ \contains , \requires , \refines , \satisfies, \conflicts \}
\end{gather}

\item Reasoning about \textit{requires} relation:
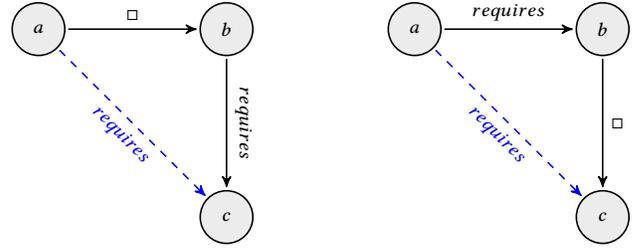
\begin{figure}[H]
\centering
      \begin{tikzpicture}[modal,node distance=2.5cm,world/.append style={minimum size=0.7cm}]

        \node (A) [world] {$a$};
        \node (B) [world, right of=A] {$b$};
        \node (C) [world, below of=B] {$c$};
        \node (D) [world, right of=B] {$a$};
        \node (E) [world, right of=D] {$b$};
        \node (F) [world, below of=E] {$c$};

        \path [->]
                (A) edge node {$\rsquare$} (B)
                (B) edge node[sloped,anchor=south] {$requires$} (C)
                (D) edge node {$requires$} (E)
                (E) edge node {$\rsquare$} (F);

        \path [->] [blue, dashed]
            (A) edge node[sloped,anchor=north] {$requires$} (C)
            (D) edge node[sloped,anchor=north] {$requires$} (F);
      \end{tikzpicture}
\caption{Inferring "\textit{requires}" with "\textit{contains}" and "\textit{refines}"}
\end{figure}

\begin{gather}\label{axioms_requires} 
\vdash \forall a, b, c \in A \mid (a,b) \in \rsquare \wedge\, (b,c) \in \require \to (a,c) \in \require \\
\vdash \forall a, b, c \in A \mid (a,b) \in \require \wedge\, (b,c) \in \rsquare \to (a,c) \in \require \\
\notag\mbox{\quad where \:} \, \square \in \{ \requires, \refines , \contains \}
\end{gather}

\item Reasoning about \textit{satisfies} relation:
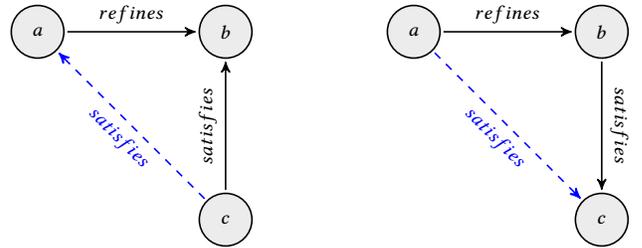
\begin{figure}[H]
\centering
    \begin{tikzpicture}[modal,node distance=2.5cm,world/.append style={minimum size=0.7cm}]
        \node (A) [world] {$a$};
        \node (B) [world, right of=A] {$b$};
        \node (C) [world, below of=B] {$c$};
        \node (D) [world, right of=B] {$a$};
        \node (E) [world, right of=D] {$b$};
        \node (F) [world, below of=E] {$c$};

        \path [->]
            (A) edge node {$refines$} (B)
            (C) edge node[sloped,anchor=south] {$satisfies$} (B)
            (D) edge node[above] {$refines$} (E)
            (E) edge node[sloped,anchor=south] {$satisfies$} (F);

        \path [->] [blue, dashed]
            (C) edge node[sloped,anchor=north] {$satisfies$} (A)
            (D) edge node[sloped,anchor=north] {$satisfies$} (F);
    \end{tikzpicture}
\caption{Patterns to infer "\textit{satisfies}" traces}
\label{fig:reasoning_satisfies}
\end{figure}

\begin{gather}\label{axioms_satisfies} 
\vdash \forall a, b, c \in A \mid (a,b) \in \rsquare \wedge\, (c,b) \in \satisfy \to (c,a) \in \satisfy \\
\vdash \forall a, b, c \in A \mid (a,b) \in \rsquare \wedge\, (b,c) \in \satisfy \to (a,c) \in \satisfy \\
\notag\mbox{\quad where \:} \, \square = \refines
\end{gather}

\item The following axiom schema is being used for generating \textit{conflicts} relation.

\begin{figure}[H]
\centering
      \begin{tikzpicture}[modal,node distance=2.5cm,world/.append style={minimum size=0.7cm}]

        \node (A) [world] {$a$};
        \node (B) [world, right of=A] {$b$};
        \node (C) [world, below of=B] {$c$};

        \path [->]  (A) edge node {$ a \rsquare b$} (B);
        \path [<->] (B) edge node {$conflicts_1$} (C);

        \path [<->] (A) edge[blue, dashed] node[sloped, anchor=north] {$conflicts_2$} (C) ;

      \end{tikzpicture}
\caption{Inferring "\textit{conflicts}"}
\end{figure}
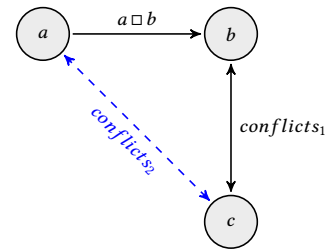

\begin{gather}\label{axioms_conflicts}
\hspace*{-1.1em}\vdash \forall a, b, c \in A \mid (a,b) \in \square \wedge (b,c) \in \conflicts \to (a,c) \in \conflicts \\
\hspace*{-1em}\vdash \forall a \in A \mid (a,a) \in \conflicts \\
\notag\mbox{\quad where \:} \square \in \{ \requires , \refines , \contains \} \mbox{ and \:} \triangle = \conflicts
\end{gather}

\item In the following axiom schema, \eqref{transitivity} is used for reasoning new traces, whereas \eqref{anti-symmetry} and \eqref{irreflexivity} are used to check consistency.
\begin{gather}
\vdash \forall a, b, c \in A \mid (a,b) \in \square \wedge (b,c) \in \square \to (a,c) \in \square  , \label{transitivity}  \\
\vdash \forall a, b \in A \mid (a,b) \in \square \wedge (b,a) \in \square \to a = b , \label{anti-symmetry}  \\
\vdash \forall a \in A \mid (a,a) \notin \square,  \label{irreflexivity} \\
\notag\mbox{where \:}  \square \in \{ \contains , \requires , \refines , \satisfies \} 
\end{gather}

\item Consistency of \textit{contains} Relation, which is left-unique (injective relation)
\begin{gather}
\vdash \forall a, a', b \in A \mid (a,b) \in \square \wedge (a',b) \in \square \to a = a' \\
\notag\mbox{\quad where \:} \, \square = \contains
\end{gather}
    
\item For instance, in addition to the previously defined axioms using the following axiom schema \eqref{consistency}, the system detects an inconsistency between two requirements, $r_{60}$ and $r_{59}$ as shown in Fig.~\ref{fig:consistency_checking}. 
\begin{gather}
\forall a, b \in A \mid (a,b) \in \square \to (a,b) \notin \triangle \wedge (b,a) \notin \triangle, \mbox{\, where} \label{consistency} \\
\notag \mbox{for each } \square \in  \{ \requires , \refines , \satisfies , \contains , \conflicts \} \\
\notag ( \requires \cup \refines \cup \satisfies \cup \contains \cup \conflicts) \setminus \square \mapsto \triangle 
\end{gather}

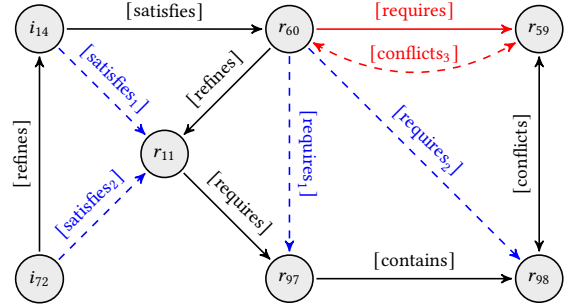
\begin{figure}[H]
\centering
    \begin{tikzpicture}[modal,node distance=1.7cm,world/.append style={minimum size=0.6cm}]
        \node[world] (11) [] {$r_{11}$};
        \node[world] (60) [above right=of 11] {$r_{60}$};
        \node[ghost] (60_59) [below right=of 60] {$r_{60}$};
        \node[world] (97) [below right=of 11] {$r_{97}$};
        \node[world] (59) [above right=of 60_59] {$r_{59}$};
        \node[world] (98) [below right=of 60_59] {$r_{98}$};
        
        \node[world] (14)[above left=of 11] {$i_{14}$};
        \path[->] (14) edge[] node[sloped,anchor=south] {$[\satisfies]$}  (60);
        \path[->] (14) edge[dashed, color=blue] node[sloped,anchor=south] {$[\satisfies_1]$}(11);

        \node[world] (72)[below left=of 11] {$i_{72}$};
        \path[->] (72) edge[] node[sloped,anchor=south] {$[\refines]$}  (14);
        \path[->] (72) edge[dashed, color=blue] node[sloped,anchor=south] {$[\satisfies_2]$}(11);

        \path[->] (60) edge[] node[sloped,anchor=south] {$[\refines]$}  (11);
        \path[->] (60) edge[dashed,color=blue] node[sloped,anchor=south] {$[\requires_1]$} (97);
        \path[->] (11) edge[] node[sloped,anchor=south]{$[\requires]$} (97);
        \path[->] (60) edge[color=red] node[sloped,anchor=south] {$[\requires]$}(59);
        \path[->] (60) edge[dashed, color=blue] node[sloped,anchor=south] {$[\requires_2]$}(98);
        \path[<->] (60) edge[bend right, dashed, color=red] node[sloped,anchor=south] {$[\conflicts_3]$}(59);
        \path[->] (97) edge[] node[sloped,anchor=south] {$[\contains]$}(98);
        \path[<->] (98) edge[] node[sloped,anchor=south] {$[\conflicts]$}(59);
    \end{tikzpicture}
\caption{Example Inferred and Inconsistent Traces} \label{fig:consistency_checking_1}
\end{figure}

\end{enumerate}

\end{document}